\documentclass[aps,prb,twocolumn,superscriptaddress,showpacs]{revtex4-1}
\usepackage[pdftex]{graphicx}
\usepackage{amsmath}
\usepackage{amssymb}

\begin{document}

\title{The effective mass of bilayer graphene: electron-hole asymmetry and electron-electron interaction}

\author{K. Zou}
\affiliation{Department of Physics, The Pennsylvania State
University, University Park, PA 16802}
\author{X. Hong}
\affiliation{Department of Physics, The Pennsylvania State
University, University Park, PA 16802}
\affiliation{Department of Physics and Astronomy, University of Nebraska-Lincoln, Lincoln, NE 68588}
\author{J. Zhu}
\affiliation{Department of Physics, The Pennsylvania State University, University Park, PA 16802}
\affiliation{The Materials Research Institute, The Pennsylvania State University, University Park, PA 16802}

%\date{\today}

\begin{abstract}
We report precision measurements of the effective mass $m^{\ast}$ in high-quality bilayer graphene using the temperature dependence of the Shubnikov-de Haas oscillations. In the density range of 0.7 x 10$^{12}$/cm$^{2}$ $<$ $n$ $<$ 4.1 x 10$^{12}$/cm$^{2}$, both the hole mass $m^{\ast}_{\mathrm{h}}$ and the electron mass $m^{\ast}_{\mathrm{e}}$ increase with increasing density, demonstrating the hyperbolic nature of the bands. The hole mass $m^{\ast}_{\mathrm{h}}$ is approximately 20-30\% larger than the electron mass $m^{\ast}_{\mathrm{e}}$. Tight-binding calculations provide a good description of the electron-hole asymmetry and yield an accurate measure of the inter-layer hopping parameter $v_{4} = 0.063$. Both $m^{\ast}_{\mathrm{h}}$ and $m^{\ast}_{\mathrm{e}}$ are suppressed compared to single-particle values, suggesting electron-electron interaction induced renormalization of the band structure of bilayer graphene.

\end{abstract}

\maketitle
\section{INTRODUCTION}

Bilayer graphene may be a technologically important material in electronics and photonics due to its tunable band gap. The fundamental property that underpins such applications-its band structure-has been the subject of many recent theoretical~\cite{Mccann2006,Mucha-Kruczynski2010,Nilsson2008,Gava2009} and experimental studies using angle-resolved photoemission spectroscopy~\cite{Ohta2006}, infrared and Raman measurements~\cite{Malard2007,Zhang2008,Kuzmenko2009}, cyclotron mass~\cite{Castro2007} and compressibility measurements~\cite{Henriksen2010,Young2010}. On a single particle level, the band structure of bilayer is thought to be well described by a tight-binding Hamiltonian~\cite{Mucha-Kruczynski2010,Nilsson2008} with a few leading order Slonczewski-Weiss-McClure parameters, i.e., $\gamma_{0}$, $\gamma_{1}$, $\gamma_{3}$ and $\gamma_{4}$. Experimental knowledge of these hopping parameters in bilayer varies, with $\gamma_{1} = 0.40$ eV fairly accurately known~\cite{Zhang2008, Kuzmenko2009} and the rest to a much less degree. For example, experimental values of $\gamma_{4}$, which controls the band asymmetry, range from 0.11-0.19 eV~\cite{Malard2007,Zhang2008,Kuzmenko2009, Henriksen2010}.

Meanwhile, electron-electron interactions in single-layer and bilayer graphene are predicted to be strong and peculiar. Interesting collective states emerge in a magnetic field~\cite{Zhao2010,Martin2010}. The many-body corrections to Fermi liquid parameters such as the compressibility $\kappa$ and the effective mass $m^{\ast}$ are expected to be sunstantial already at currently accessible densities~\cite{Kusminskiy2008,Kusminskiy2009,Borghi2009,Borghi2010,Sensarma2011}. These renormalization effects are related to, but also quantitatively different from those observed in conventional two-dimensional electron gases (2DEGs)~\cite{Giuliani2005, Tan2005}, due to the chirality of single and bilayer graphene~\cite{Borghi2009}. For example, instead of an enhancement~\cite{Tan2005}, the effective mass of bilayer graphene is predicted to be increasingly suppressed at lower carrier densities~\cite{Borghi2009}.  No experimental evidence of such renormalization effect has been reported so far.

In this work, we report the measurements of the effective mass $m^{\ast}$ in bilayer graphene samples for a wide range of carrier densities using high-quality Shubnikov-de Haas (SdH) oscillations. The inter-layer hopping parameter $\gamma_{4}$ is determined to be $\gamma_{4}$ = 0.063(1)$\gamma_{0}$ with the highest accuracy reported so far. The magnitude and density dependence of $m^{\ast}$ deviate from tight-bind calculations, providing evidence for electron-electron interaction induced band renormalization.

\section{SAMPLE PREPARATION}
Bilayer graphene flakes are exfoliated onto 290 nm SiO$_{2}$/Si wafers from highly ordered pyrolytic graphite and identified by optical microscope and Raman spectra. They are further confirmed by their quantum Hall sequence. Conventional electron-beam lithography is used to pattern the flakes into Hall bars.

\section{MEASUREMENTS}
Transport measurements are carried out in a He$^{4}$ system using standard low-frequency lock-in technique. The field effect mobility $\mu_{\mathrm{FE}} = (1/e)(\mathrm{d}\sigma/\mathrm{d}n)$ of our pristine bilayer graphene ranges from 3 000-12 000 cm$^{2}$/Vs. Data from two samples (A and B) are presented in this paper.

\begin{figure}
\includegraphics[angle=0,width=2.6in]{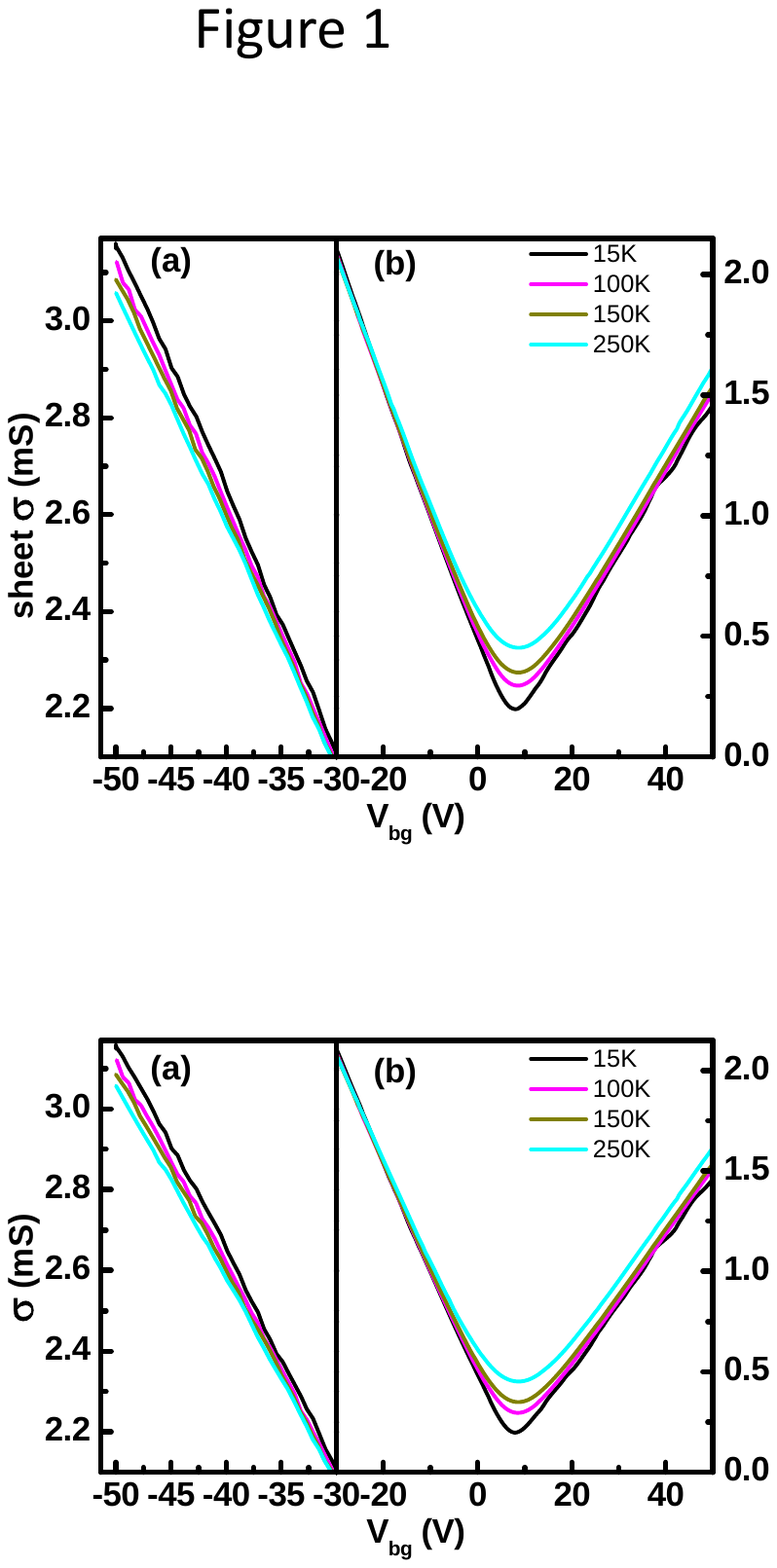}
\vspace{-0.2in}
 \caption[]{(color online) (a) and (b) The sheet conductance $\sigma(V_{\mathrm{bg}})$ of sample A. From top to bottom: $T$= 15, 100, 150, 250 K in (a) and in reverse order in (b). The charge neutrality point is at $V_{\mathrm{bg}}$= 7 V.
\label{sigT}}
\vspace{-0.2in}
\end{figure}

In Fig.~\ref{sigT}, we plot the sheet conductance $\sigma$ vs. the back-gate voltage $V_{\mathrm{bg}}$ of sample A at selected temperatures between 15 K--250 K. At 15 K, the mobility $\mu_{\mathrm{FE}}$ of sample A is approximately 4 800 cm$^{2}$/Vs for holes and 3 100 cm$^{2}$/Vs for electrons. Sample B has a higher mobility of 6 300 cm$^{2}$/Vs for holes and 6 800 cm$^{2}$/Vs for electrons. The conductance of bilayer graphene samples shows a variety of temperature dependence, depending on the carrier density and mobility. Near the charge neutrality point, all our samples show an insulating-like $T$-dependence ($\mathrm{d}\sigma/\mathrm{d}T$ $>$ 0), as shown in Fig.~\ref{sigT}. This behavior is due to the thermal excitation of carriers out of electron-hole puddles, as demonstrated in Ref.~\onlinecite{Zou2010b}. As the carrier density increases, $\mathrm{d}\sigma/\mathrm{d}T$ eventually becomes negative (metallic) in the highest-quality samples. This trend is shown by the hole branch in Fig.~\ref{sigT}(a) and (b), where the crossover density is approximately $n_{\mathrm{h}}$ = 2.1 x 10$^{12}$/cm$^{2}$ . For samples with lower mobilities, the insulating-like $T$-dependence persists to high densities, an example of which is given by the electron branch in Fig.~\ref{sigT}(b).

This complex behavior is in contrast to that of single-layer graphene, where a metallic-like temperature dependence dominates over a wide range of densities due to phonon scattering~\cite{Chen2009,Zhu2009,Zou2010l}. The qualitative features of our data are consistent with the model proposed in Ref.~\onlinecite{Hwang2010}, where $\sigma(T)$ combines metallic and insulating trends arising from the conduction of the majority and minority carriers respectively. The true metallic $T$-dependence of a bilayer graphene two-dimensional electron gas emerges only in high quality samples and/or at high carrier densities. In Fig.~\ref{sigT}, the different $T$-dependence of the two carriers in the same sample points to an intrinsic electron-hole (e-h) asymmetry of bilayer graphene, which we further examine below.

\begin{figure}
\includegraphics[angle=0,width=2.8in]{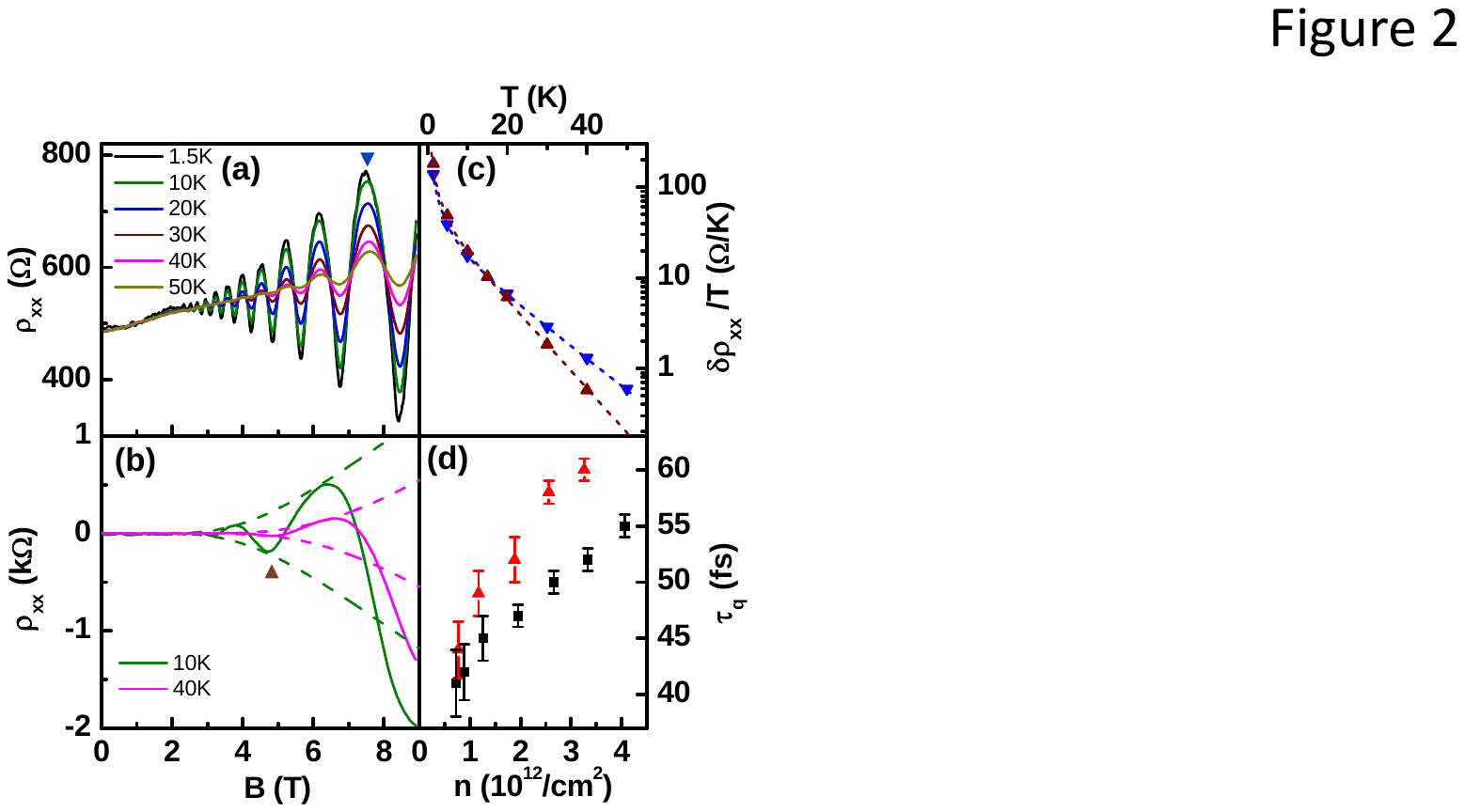}
\vspace{-0.1in}
 \caption[]{(color online) (a) SdH oscillations $\rho_{\mathrm{xx}}(B)$ at $T$ = 1.5-50 K for $n_{\mathrm{e}}$ = 3.26x10$^{12}$/cm$^{2}$. (b) $\rho_{\mathrm{xx}}(B)$ at $T$ = 10 K and 40 K for $n_{\mathrm{h}}$ = 0.89 x 10$^{12}$/cm$^{2}$. Dashed lines are fittings with $\tau_{\mathrm{q}}$ = 42 fs and $m^{\ast}_{\mathrm{h}}$ = 0.036$m_{\mathrm{e}}$. A smooth background has been subtracted. (c) $\delta\rho_{\mathrm{xx}}/T$ vs. $T$ in a semi-log plot for $n_{\mathrm{e}}$ = 3.26 x 10$^{12}$/cm$^{2}$ at $B$ = 7.53 T (down triangle in (a)) and for $n_{\mathrm{h}}$ = 0.89x10$^{12}$/cm$^{2}$ at $B$ = 4.70 T (up triangle in (b)). The symbols correlate. Dashed lines are fittings with $m^{\ast}_{\mathrm{e}}$ = 0.041$m_{\mathrm{e}}$ (down triangles) and $m^{\ast}_{\mathrm{h}}$ = 0.036$m_{\mathrm{e}}$ (up triangles).(d) $\tau_{\mathrm{q}}$ vs. $n$ for electrons (red triangles) and holes (black squares). All data in (a)-(d) are from sample A.
\label{osc}}
\vspace{-0.3in}
\end{figure}

To probe the band structure of bilayer graphene, we measure the effective mass $m^{\ast}$ as a function of the carrier density using SdH oscillations. This technique is well established in 2DEGs but require high-quality oscillations to reliably extract $m^{\ast}$. Figure~\ref{osc}(a) shows the SdH oscillations $\rho_{\mathrm{xx}}(B)$ of sample A at a high electron density $n_{\mathrm{e}}$ = 3.26 x 10$^{12}$/cm$^{2}$ and varying temperatures. The oscillations have an early onset, appear sinusoidal and free of beating. Its amplitude $\delta\rho_{\mathrm{xx}}$ is given by~\cite{Hong2009}:
\begin{equation}
\frac{\delta\rho_{\mathrm{xx}}}{\rho_{0}}=4\gamma_{\mathrm{th}}\mathrm{exp}(-\frac{\pi}{\omega_{\mathrm{c}}\tau_{q}});\gamma_{\mathrm{th}}=\frac{2\pi^{2}k_{\mathrm{B}}T/\hbar\omega_{\mathrm{c}}}{\mathrm{sinh}(2\pi^{2}k_{\mathrm{B}}T/\hbar\omega_{\mathrm{c}})}
\label{equ1}
\vspace{-0.1in}
\end{equation}
where $\omega_{\mathrm{c}}=eB/m^{\ast}$ is the cyclotron frequency, $\tau_{\mathrm{q}}$ is the quantum scattering time and $\gamma_{\mathrm{th}}$ the thermal factor.

As shown in Fig.~\ref{osc}(a), $\delta\rho_{\mathrm{xx}}$ increases with increasing $B$ and decreasing $T$ . Its $T$-dependence provides a direct measure of $m^{\ast}$ whereas the $B$-dependence is controlled by both $m^{\ast}$ and $\tau_{\mathrm{q}}$. At each carrier density, the low-field $\delta\rho_{\mathrm{xx}}(T, B)$ data, i.e., before the onset of the quantum Hall effect, is fit to Eq.~\ref{equ1} with two fitting parameters $m^{\ast}$ and $\tau_{\mathrm{q}}$. The simultaneous fitting of $m^{\ast}$ and $\tau_{\mathrm{q}}$ allows us to accurately determine $\delta\rho_{\mathrm{xx}}$, especially at low carrier densities, where the oscillations are few and a linear interpolation between peaks, as commonly done in the literature~\cite{Tan2005}, cannot give the correct amplitude of $\delta\rho_{\mathrm{xx}}$.  Figure~\ref{osc}(b) shows $\rho_{\mathrm{xx}}(B)$ data at T=10 K and 40 K for a low hole density $n_{\mathrm{h}}$ = 0.89 x 10$^{12}$/cm$^{2}$. Fittings to Eq.~\ref{equ1} are shown as dashed lines. Only the right values of $m^{\ast}$ and $\tau_{\mathrm{q}}$ can fit both the $B$-dependence and $T$-dependence of $\delta\rho_{\mathrm{xx}}$ simultaneously.

 In Fig.~\ref{osc}(c), we plot two examples of the measured $\delta\rho_{\mathrm{xx}}/T$ vs. $T$ in a semi-log plot for the two positions marked in Figs.~\ref{osc}(a) and (b) with down and up triangles respectively. Dashed lines are fittings generated with $m^{\ast}_{\mathrm{e}}$ = 0.041$m_{\mathrm{e}}$ and $m^{\ast}_{\mathrm{h}}$ = 0.036$m_{\mathrm{e}}$ respectively, where $m_{\mathrm{e}}$ is the electron rest mass. They both fit very well. Overall, Eq.~\ref{equ1} provides an excellent description of the $\delta\rho_{\mathrm{xx}}(T, B)$ data in the entire density range studied, with the uncertainty of $m^{\ast}$ increasing from 0.0001$m_{\mathrm{e}}$ to 0.0015$m_{\mathrm{e}}$ from high to low densities. The global fitting procedure also ensures that the extracted $m^{\ast}$ is filling factor independent and therefore represents the $B$ = 0 limit, i.e., the band structure mass. This $m^{\ast}$ is not directly comparable to $m^{\ast}$ determined from cyclotron resonance measurements~\cite{Henriksen2008cr}, as Coulomb interaction may manifest differently in these two cases~\cite{Shizuya2010,*Zhang2010}. A good illustration of this situation is the parabolic band material GaAs, where $m^{\ast}$ determined from SdH oscillations embodies electron-electron (e-e) interaction~\cite{Tan2005} while its effect is forbidden in cyclotron resonance measurements by the Kohn theorem~\cite{Kohn1961}.

Using this method, we have determined $m^{\ast}$ and $\tau_{\mathrm{q}}$ for samples A and B in the density range of 0.7 $<$ $n$ $<$ 4.1 x 10$^{12}$/cm$^{2}$ for electrons and holes. Both samples show oscillations of equally high quality and comparable $\tau_{\mathrm{q}}$ . Figure~\ref{osc}(d) plots $\tau_{\mathrm{q}}(n)$ of sample A for both carriers. Overall, $\tau_{\mathrm{q}}$ increases with increasing density, ranging from 41 to 60 fs. These values correspond to a disorder broadening $\Gamma=\hbar/2\tau_{\mathrm{q}}$ = 5.5-8.0 meV, which are similar to high-quality single and bilayer graphene samples~\cite{Zhang2008,Hong2009}.

The results of $m^{\ast}$ of samples A and B as a function of the carrier density $n$ are plotted in Fig.~\ref{mass}(a). The error bars represent uncertainties obtained from fittings similar to those shown in Fig.~\ref{osc}(c). The two samples agree very well with each other. In the density range studied, both $m^{\ast}_{\mathrm{e}}$ and $m^{\ast}_{\mathrm{h}}$ increase with increasing $n$, indicating the non-parabolic nature of the bands. This observation agrees with the compressibility measurements of Refs.~\onlinecite{Henriksen2010,Young2010} and is also consistent with the observation of a constant $m^{\ast}$ at yet lower densities~\cite{Martin2010}. The ratio of $m^{\ast}_{\mathrm{h}}$/$m^{\ast}_{\mathrm{e}}$ is about 1.2-1.3, demonstrating a pronounced electron-hole asymmetry.

\section{ANALYSIS AND DISCUSSION}
The above measurements of $m^{\ast}$ provide an accurate means of determining the band structure of bilayer graphene and investigating the effect of e-e interaction. In the following analysis, we employ a tight-binding Hamiltonian following the notations of Refs.~\onlinecite{Nilsson2008} and~\onlinecite{Zhang2008}:
\begin{equation}
H=\left( \begin{array}{cccc}
V(n)/2+\Delta&\phi&\gamma_{1}&-v_{4}\phi^{\ast}\\
\phi^{\ast}&V(n)/2&-v_{4}\phi^{\ast}&v_{3}\phi\\
\gamma_{1}&-v_{4}\phi&-V(n)/2+\Delta&\phi^{\ast}\\
-v_{4}\phi\ & v_{3}\phi^{\ast}&\phi&-V(n)/2
\end{array} \right).
\label{equ2}
\end{equation}

Equation~\ref{equ2} is written in the basis of the four sublattices ($\Psi_{\mathrm{A1}}, \Psi_{\mathrm{B1}}, \Psi_{\mathrm{A2}}, \Psi_{\mathrm{B2}}$), where A1, A2 are the two stacked sublattices in layer 1 and 2 respectively. The nearest neighbor in-plane (A1-B1) hopping integral $\gamma_{0}$ is included in $\phi=\gamma_{0}(3/2k_{\mathrm{y}}a-\mathrm{i}3/2k_{\mathrm{x}}a)=\hbar v_{\mathrm{F}}(k_{\mathrm{y}}-\mathrm{i}k_{\mathrm{x}})$,  where $a$ = 1.42${\mathrm{\AA}}$ is the carbon-carbon distance and the momentum vector $(k_{\mathrm{x}}, k_{\mathrm{y}})$ originates from the K (K$^{\prime}$) point of the Brillouin zone. The Fermi velocity $v_{\mathrm{F}}=(3/2)\gamma_{0}a/\hbar$. $\gamma_{1}, v_{3}=\gamma_{3}/\gamma_{0}$ and $v_{4}=\gamma_{4}/\gamma_{0}$ represent the hopping integrals between two inter-layer sublattices A1-A2, B1-B2, and A1-B2 respectively. $\gamma_{1}$ gives rise to the band splitting, $\gamma_{3}$ leads to trigonal warping of the Fermi surface and $\gamma_{4}$ controls electron-hole asymmetry. $\Delta$ is the on-site energy difference of A1 and B1, due to their stacking difference. $V(n)$ is the potential difference between the two layers and varies with the carrier density~\cite{Zhang2008}. The eigenvalues of Eq.~\ref{equ2} produce the four low-energy bands of bilayer graphene. Out of the four bands, the two higher energy electron and hole bands are neglected here since they are far above the Fermi level of our density range, $E_{\mathrm{F}} \sim$ 30-120 meV. The effective mass $m^{\ast}$ of the lower bands is given by:
\begin{equation}
m^{\ast}=\frac{\hbar^{2}}{2\pi}\frac{\mathrm{d}A(E)}{\mathrm{d}E}\mid_{E=E_{\mathrm{F}}},
\label{equ3}
\vspace{-0.1in}
\end{equation}
where $A(E)$ is the $k$-space area enclosed by  the contour of constant energy $E$. For $\gamma_{3}$ = 0, the contour is circular and Eq.~\ref{equ3} is simplified to $m^{\ast}=\hbar^{2}k/(\mathrm{d}E(k)/\mathrm{d}k$).

We diagonalize Eq.~\ref{equ2} and numerically compute $m^{\ast}$ using Eq.~\ref{equ3}. The effect of each parameter in Eq.~\ref{equ2} on $m^{\ast}$ is summarized in table~\ref{table1}, where +(-) means an increase of the parameter will increase (decrease) the value of $m^{\ast}$.
\begin{table}
\caption[]{The effect of tight-binding parameters on $m^{\ast}$ and their values.}
\begin{ruledtabular}
\begin{tabular}{|c|c|c|c|c|c|c|}
   & $\gamma_{0}$ & $\gamma_{1}$ & $\gamma_{3}$ & $V(n)$ & $v_{4}$ & $\Delta$\\
\hline
$m_{\mathrm{h}}^{\ast}$ & - & + & + & + & + & +\\
\hline
$m_{\mathrm{e}}^{\ast}$ & - & + & + & + & - & -\\
\hline
value (eV) & 3.43(1) & 0.40$^{\dag}$ &0&& 0.063(1) & 0.018$^{\dag}$ \\
\end{tabular}
\end{ruledtabular}
\label{table1}
\dag.Reference~\onlinecite{Zhang2008}.
\vspace{-0.3in}
\end{table}

In our calculations, the inter-layer B1-B2 hopping energy $\gamma_{3}$ is set to zero due to its negligible effect in the density range considered here (see Appendix A for details). The gate voltage-induced $V(n)$ is calculated following Eqs.(7)-(13) in Ref.~\onlinecite{Zhang2008}, using self-consistent screening and including the small initial doping of our samples. Both $V(n)$ and the initial doping produce minute corrections to $m^{\ast}$ in the density range studied (see Appendix B for details). Consequently, the overall magnitude of $m^{\ast}_{\mathrm{h}}$ and $m^{\ast}_{\mathrm{e}}$ and their density dependence are predominantly controlled by $\gamma_{0}$ and $\gamma_{1}$. In the literature, $\gamma_{1}$ is found to be 0.38-0.40 eV by infrared measurements~\cite{Zhang2008, Kuzmenko2009}. Most of our fittings use $\gamma_1$=0.40 eV. Alternative scenarios are also considered in the discussion of electron-electron interaction effect and further explored in Appendix C. The difference between $m^{\ast}_{\mathrm{h}}$ and $m^{\ast}_{\mathrm{e}}$ is controlled by $v_4$ and $\Delta$. We fix $\Delta$=0.018 eV in our calculations. A 10$\%$ variation of $\Delta$ among literature values~\cite{Zhang2008,Kuzmenko2009} leads to a change smaller than 2$\%$ in $v_4$, which is comparable to its estimated uncertainty.
% remove appendix A
%In our calculations, we set $\gamma_{1}$=0.40 eV and $\Delta$=0.018 eV, using results from infrared spectroscopy~\cite{Zhang2008,Kuzmenko2009}.  The 10$\%$ variation of $\Delta$ in the literature leads to a change smaller than 1$\%$ in $v_{4}$, which is smaller than its estimated uncertainty. The inter-layer B1-B2 hopping energy $\gamma_{3}$ leads to trigonal warping of the Fermi surface~\cite{Nilsson2008}. In the density range of our experiment, even the largest $\gamma_{3}$ = 0.38 eV found in the literature~\cite{Ohta2006,Kuzmenko2009,Malard2007} results only a minute increase of $m^{\ast}$, comparable to the smallest error bar. We therefore set $\gamma_{3}$ = 0 forthward.

%The gate voltage-induced $V(n)$ is calculated following Eqs.(7)-(13) in Ref.~\cite{Zhang2008}, which agrees well with optical measurements~\cite{Zhang2009}. The calculation includes a small initial hole doping of $V_{\mathrm{bg}}$ = +7 V for sample A and $V_{\mathrm{bg}}$ = +17 V for sample B, likely due to water adsorbed on the top layer. It also includes the quantum level broadening $\Gamma$ calculated from $\tau_{\mathrm{q}}$. The calculated $V(n)$ varies from 2-70 meV. The opening of a band gap increases both $m^{\ast}_{\mathrm{h}}$ and $m^{\ast}_{\mathrm{e}}$ slightly. The effect is the largest at $V$ = 70 meV, where both $m^{\ast}_{\mathrm{e}}$ and $m^{\ast}_{\mathrm{h}}$ increase by ~0.002$m_{\mathrm{e}}$, which is $\sim5\%$ of measured $m^{\ast}$.

\begin{figure}
\includegraphics[angle=0,width=3.2in]{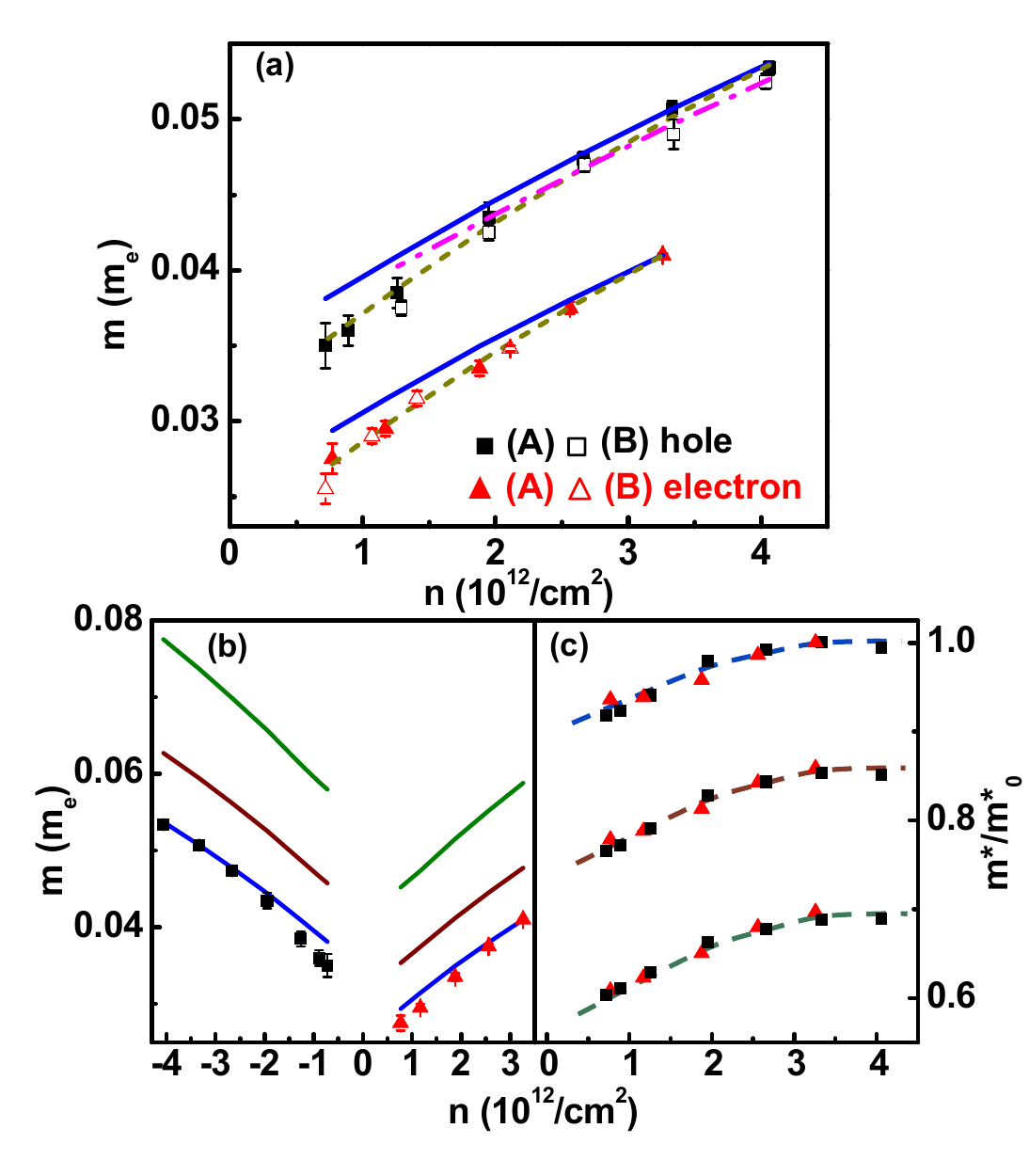}
\vspace{-0.2in}
 \caption[]{(color online)(a) Measured $m^{\ast}_{\mathrm{h}}$ and $m^{\ast}_{\mathrm{e}}$ vs. $n$ for samples A and B. The symbols are indicated in (a) and used in (a)-(c). Solid blue lines are fittings to sample A with $\gamma_{\mathrm{0}}$ = 3.419 eV, $\gamma_{\mathrm{1}}$ = 0.40 eV and $v_4$=0.063. The magenta dash-dotted line is a fitting to sample B with $\gamma_{\mathrm{0}}$ = 3.447 eV, $\gamma_{\mathrm{1}}$ = 0.40 eV and $v_4$=0.063. The yellow short dashed lines correspond to $\gamma_{\mathrm{0}}$ = 3.167 eV, $\gamma_{\mathrm{1}}$ = 0.30 eV and $v_4$=0.063. (b) A comparison of measured $m^{\ast}$ and calculated $m^{\ast}_{\mathrm{0}}$ for sample A. From top to bottom: $\gamma_{\mathrm{0}}$ = 2.72 (olive), 3.09 (wine), and 3.42 eV(blue). $\gamma_{\mathrm{1}}$ = 0.40 eV, $v_4$=0.063 and $\Delta$ = 0.018 eV for all traces. (c) The ratio $m^{\ast}$/$m^{\ast}_{\mathrm{0}}$ vs. $n$ for sample A. From top to bottom: $\gamma_{\mathrm{0}}$ = 3.42, 3.09, and 2.72 eV. Dashed lines are guide to the eye.
\label{mass}}
\vspace{-0.3in}
\end{figure}

Fitting to $m^{\ast}_{\mathrm{h}}$ and $m^{\ast}_{\mathrm{e}}$ simultaneously allows us to determine the remaining adjustable parameters, $\gamma_{0}$ and $v_{4}$. Fittings to both samples are given in Fig.~\ref{mass}(a). The value of $\gamma_{0}$ varies slightly from 3.419 eV in A to 3.447 eV in B, yielding an average $\gamma_{0}$=(3.43 $\pm$ 0.01) eV. This corresponds to a $v_F=1.11\times 10^6$ m/s, in agreement with previous experiments. Both samples yield $v_{4}$ = 0.063 $\pm$ 0.001. The value of $v_{4}$ is also independent of the choice of $\gamma_{1}$, as the fittings in Fig.~\ref{mass}(a) show.  This result is consistent with the range of $v_{4}\sim$0.04--0.06 obtained previously~\cite{Zhang2008,Kuzmenko2009,Malard2007,Henriksen2010}, but has a much higher precision. This accurate knowledge of electron-hole asymmetry will be important to potential electronic and optical applications of bilayer graphene.

The above fitting does not include the in-plane next-nearest neighbor hopping integral $\gamma_{\mathrm{n}}$~\cite{Mucha-Kruczynski2010}, which also contributes to the electron-hole asymmetry of $m^{\ast}$, acting in the opposite direction of $v_{4}$~\cite{note}. The value of $\gamma_{\mathrm{n}}$ is not well established. Including an additional diagonal term $-\gamma_{\mathrm{n}}\mid\phi\mid^{2}/\gamma_{1}^{2}$ in Eq.~\ref{equ2}~\cite{Mucha-Kruczynski2010}, our calculations show that the effect of $\gamma_{\mathrm{n}}$ on $v_{4}$ can be represented by an empirical relation $v_{4}$ = 0.063 + 0.037$\gamma_{\mathrm{n}}$, which can provide further correction to $v_{4}$ should the value of $\gamma_{\mathrm{n}}$ become known.

The fittings in Fig.~\ref{mass}(a) reveal an important trend of our data, i.e., the measured $m^{\ast}$ increasingly drops below the calculated $m^{\ast}$ as $n$ decreases. This trend is consistently seen for both electrons and holes and in both samples. Extensive tests show that this discrepancy between data and tight-binding calculations cannot be reconciled by varying any other parameters except for $\gamma_{1}$. A perfect fit to both high and low densities is only possible if $\gamma_{1}$ is allowed to decrease from 0.40 eV to 0.30 eV, as shown by the short-dashed lines in Fig.~\ref{mass}(a). This scenario, although appealing, is at odds with previous experimental determination of $\gamma_{1}$=0.38-0.40 eV from infrared spectroscopy~\cite{Zhang2008,Kuzmenko2009}. Alternatively, we attribute the discrepancy between measurements and tight-binding calculations to interaction induced band renormalization effect. Indeed, a recent calculation in bilayer graphene predicts a monotonic suppression of $m^{\ast}$ as a function of decreasing density~\cite{Borghi2009} and the effect is shown to be already substantial in the density range studied here. First-principles calculations also show that a more complete inclusion of e-e interaction in the form of GW corrections increases $\gamma_{0}$ from the mean-field-like value of 2.7 eV to an interaction-modified value of 3.4 eV~\cite{Trevisanutto2008,Park2009,Gava2009}. In our experiment, the suppression of m*, its density dependence, and the fitting result of $\gamma_{0}$=3.43 eV all strongly point to the renormalization effect of e-e interaction on $m^{\ast}$.

The magnitude of this effect is illustrated in Fig.~\ref{mass}(b) and (c), using sample A as an example. Here we calculate and plot three sets of $m^{\ast}_{0}$ values using $\gamma_{0}$ = 2.72, 3.09, and 3.42 eV (corresponding to $v_F=0.88, 1.0, 1.11\times 10^6$ m/s respectively). These three values are the first principle mean-field-like, intermediate, and our fitting result of $\gamma_{0}$ respectively. The other parameters are fixed at values listed in Table~\ref{table1}. Fig.~\ref{mass}(c) plots the ratio of measured m* and the calculated $m^{\ast}_{0}$, $m^{\ast}$/$m^{\ast}_{0}$ vs $n$ for each $\gamma_{0}$. The trend of decreasing $m^{\ast}$/$m^{\ast}_{0}$ with decreasing density is seen in each case, with the magnitude of the suppression depending on the input value of $\gamma_{0}$. Electrons and holes follow the same trend. When the first principle mean-field-like value of $\gamma_{0}$ = 2.72 eV is used (bottom trace in Fig.~\ref{mass}(c)), the suppression of $m^{\ast}$ is quite large, varying from 0.6 to 0.7 in the density range 0.7 $<$ $n$ $<$ 4 x 10$^{12}$/cm$^{2}$. These observations provide the first experimental indication of e-e interaction induced band renormalization effect in bilayer graphene. The quantitative input provided by our data should constrain and guide future calculations on this important subject, as the correct theory must capture both the magnitude and the density dependence of $m^{\ast}$.

\section{CONCLUSION}
To conclude, we report the measurement of the effective mass $m^{\ast}$ in bilayer graphene over a wide range of electron and hole densities. Our results demonstrate a pronounced electron-hole asymmetry, from which we accurately determine the inter-layer hopping parameter $v_{4}$ in the tight-binding description of the band structure. The measured $m^{\ast}$ is suppressed compared to single-particle predictions, indicating the possibility of interaction-induced band renormalization at play. Our results provide critical experimental input to understand the effect of electron-electron interaction in this unique two-dimensional electron system.

\begin{acknowledgments}
We are grateful for discussions with M. Cohen, V. Crespi, M. Fogler, J. Jain, ZQ. Li and M. Polini. This work is supported by NSF CAREER grant no. DpaMR-0748604 and NSF NIRT grant no. ECS-0609243. The authors acknowledge use of facilities at the PSU site of NSF NNIN.
\vspace{-0.2in}
\end{acknowledgments}

\begin{appendix}

\section{THE EFFECT OF $\gamma_{3}$ ON $m^{\ast}$}
\begin{figure}
\includegraphics[angle=0,width=2.6in]{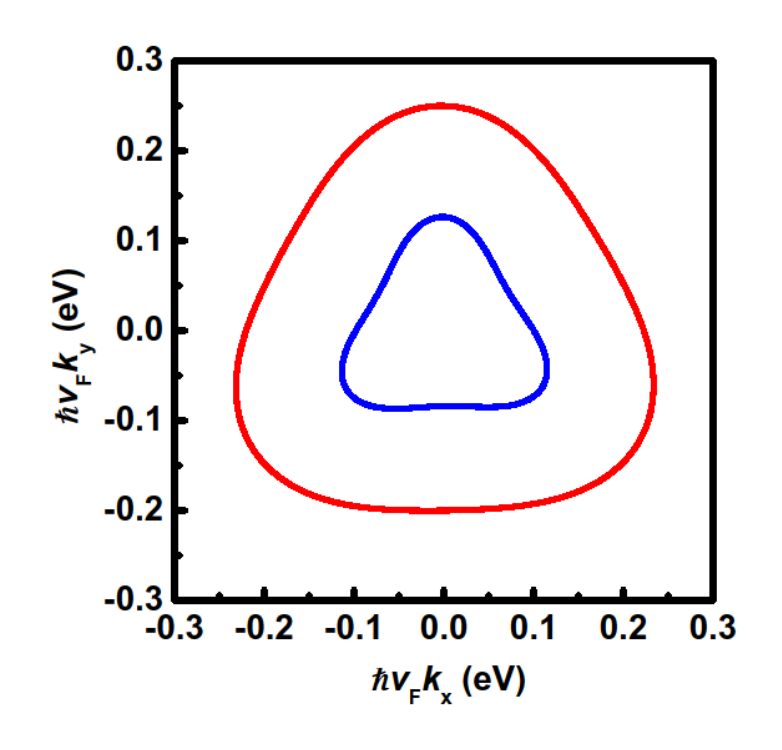}
\vspace{-0.1in}
 \caption[]{(color online) The warped Fermi surfaces in momentum space for $E_{\mathrm{F}}$ = 100 meV ($n$ $\sim$ 4 x 10$^{12}$/cm$^{2}$) (red line) and $E_{\mathrm{F}}$ = 25 meV ($n$ $\sim$ 0.7 x 10$^{12}$/cm$^{2}$) (blue line). The parameters used are $\gamma_{0}$ = 3.1 eV, $\gamma_{1}$ = 0.4 eV, $\gamma_{3}$ = 0.31 eV, $\Delta$ = 0 and $V(n)$ = 0.
\label{warp}}
\vspace{-0.2in}
\end{figure}
The inter-layer B1-B2 hopping energy $\gamma_{3}$ leads to trigonal warping of the Fermi surface~\cite{Nilsson2008}. An example is plotted in Fig.~\ref{warp} for $n$ $\sim$ 0.7 x 10$^{12}$/cm$^{2}$ and $n$ $\sim$ 4 x 10$^{12}$/cm$^{2}$. $m^{\ast}$ is calculated using Eq.~\ref{equ3} of the main text. In the density range of our experiment, even the largest $\gamma_{3}$ = 0.38 eV found in the literature~\cite{Kuzmenko2009,Ohta2006,Malard2007} results only a minute increase of $m^{\ast}$, comparable to the smallest error bar. We therefore set $\gamma_{3}$ = 0 in the calculations.

\section{THE EFFECT OF $V(n)$ AND THE INITIAL DOPING ON $m^{\ast}$}
\begin{figure}
\includegraphics[angle=0,width=2.5in]{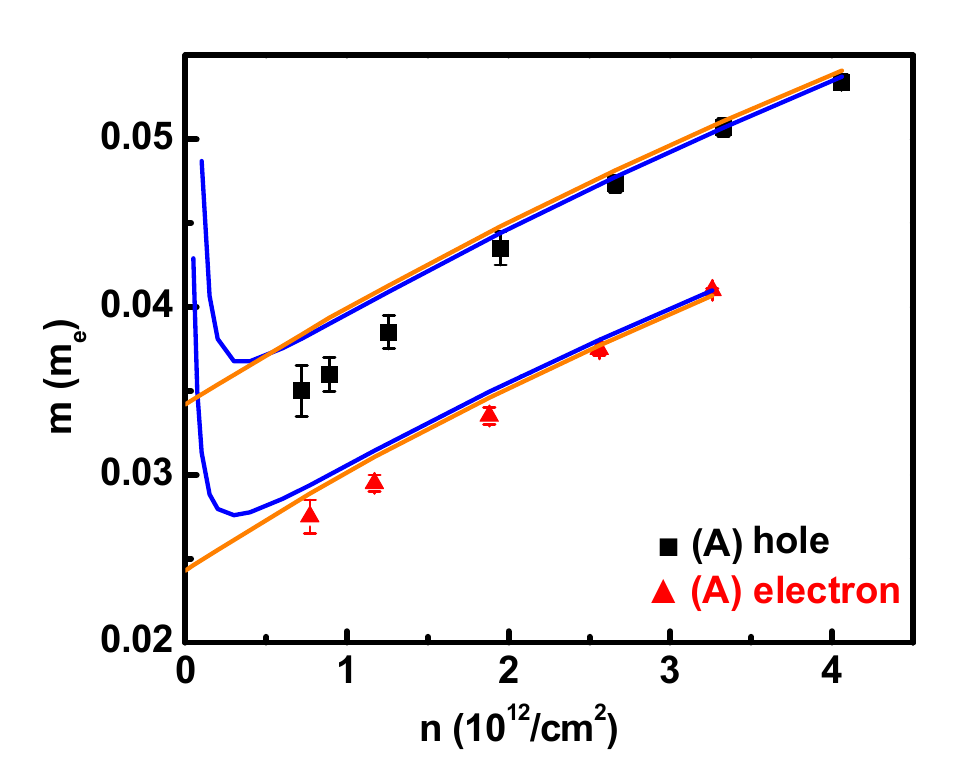}
\vspace{-0.1in}
 \caption[]{(color online) The measured (symbols) and calculated (solid lines) $m^{\ast}$ of sample A.  Blue lines correspond to an initial doping of $V_{\mathrm{bg}}$ = +7 V and orange lines correspond to zero doping. The other parameters used are $\gamma_{0}$ = 3.419 eV, $\gamma_{1}$ = 0.4 eV, $\gamma_{3}$ = 0.0 eV, $v_{4}$ = 0.063 and $\Delta$ = 0.018 eV.
\label{initialdoping}}
\vspace{-0.2in}
\end{figure}
The gate voltage-induced $V(n)$ is calculated using Eqs.(7)-(13) in Ref.~\onlinecite{Zhang2008}, which agrees well with optical measurements\cite{Zhang2009}. The calculation includes a small initial hole doping of $V_{\mathrm{bg}}$ = +7 V for sample A and $V_{\mathrm{bg}}$ = +17 V for sample B, likely due to water adsorbed on the top layer. It also includes the quantum level broadening $\Gamma$ calculated from the quantum scattering time $\tau_{\mathrm{q}}$. $V(n)$ increases with increasing $n$ and varies from 2-70 meV. As $V(n)$ increases, both $m^{\ast}_{\mathrm{h}}$ and $m^{\ast}_{\mathrm{e}}$ increase slightly compared to zero-gap cases. The effect is the largest at V = 70 meV ($n$=4.1 x 10$^{12}$/cm$^{2}$), where both $m^{\ast}_{\mathrm{h}}$ and $m^{\ast}_{\mathrm{e}}$ increase by $\sim$ 0.002 $m_{\mathrm{e}}$, which is $\sim$ 5 $\%$ of measured $m^{\ast}$.

Fig.~\ref{initialdoping} illustrates the effect of the initial chemical doping on the calculated $m^{\ast}$ in sample A. The blue lines are calculated with the measured doping of $V_{\mathrm{bg}}$ = +7 V. The orange lines are calculated with zero doping. The initial doping drastically enhances $m^{\ast}$ near the charge neutrality point but produces negligible effect in the density range studied here.

\section{THE CORRELATION BETWEEN $\gamma_{0}$ AND $\gamma_{1}$}

\begin{figure}
\includegraphics[angle=0,width=2.8in]{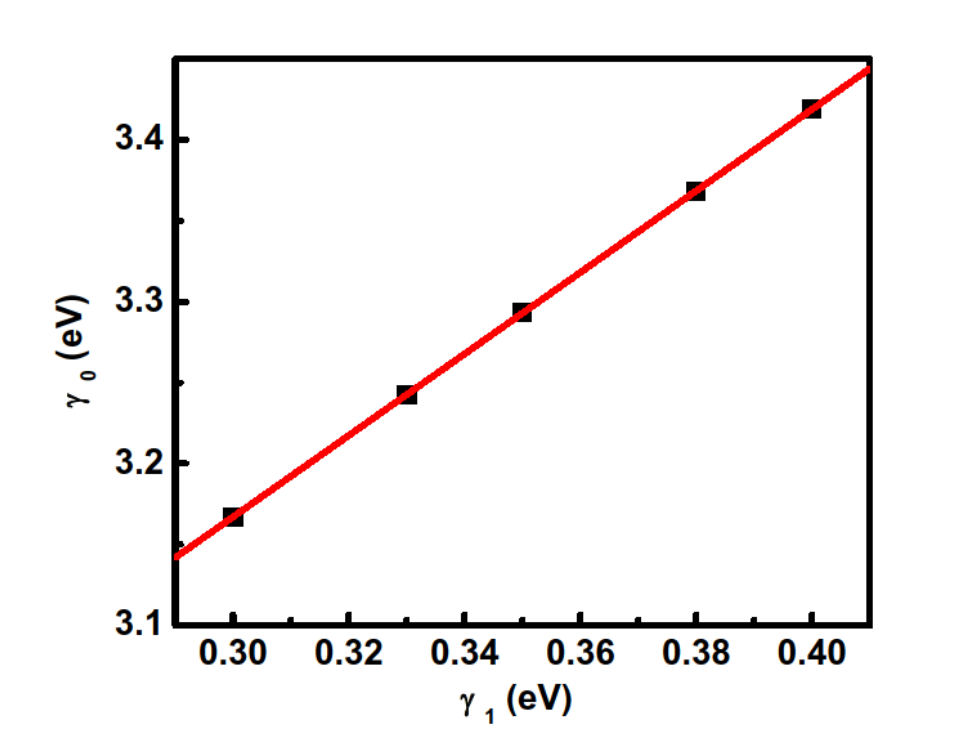}
\vspace{-0.1in}
 \caption[]{(color online) Fitting results of $\gamma_{0}$ as a function of $\gamma_{1}$. The linear fit corresponds to $\gamma_{0}$ = 2.411 + 2.52$\gamma_{1}$ .
\label{relation01}}
\vspace{-0.2in}
\end{figure}

In Fig.~\ref{mass}(a), the fitting results of $\gamma_{0}$ depend on the input parameter $\gamma_{1}$. This relationship can be described by a linear fit as shown in Fig.~\ref{relation01}. As discussed in the text, the choice of $\gamma_{1}$=0.40 eV leads to $\gamma_{0}$=3.419 eV and discrepancy between data and calculations at low densities. A decrease in $\gamma_{1}$ also decreases $\gamma_{0}$ and leads to better fit at low densities. Both high and low density data can be fit by $\gamma_{1}$=0.30 eV and $\gamma_{0}$=3.167 eV. This choice of $\gamma_{1}$ is however incompatible with the experimental range of 0.38-0.40 eV obtained from infrared absorption measurements~\cite{Zhang2008,Kuzmenko2009}.

\end{appendix}

%\bibliography{REF_MASS_07_15}
%

\end{document}